\documentclass[final,3p,times,onecolumn]{elsarticle}

\usepackage{amssymb}

\pdfoutput=1
\usepackage[table,xcdraw]{xcolor}
\usepackage{graphicx}
\usepackage{epstopdf}
\usepackage{mathrsfs}
\usepackage{amssymb}
\usepackage{verbatim}
\usepackage{color}
\usepackage{hhline}
\usepackage{stackrel}
\usepackage[pdfencoding=auto]{hyperref}
\usepackage{array,multirow}
\usepackage{colortbl}
\usepackage{hhline}
\usepackage{lipsum, babel}
\usepackage{xcolor}
\usepackage{lipsum}
\usepackage{amsmath}
\usepackage{multirow}
\usepackage{tabularx}
\usepackage{booktabs}
\usepackage{physics}
\usepackage{hyperref}
\usepackage[table,xcdraw]{xcolor}

\hypersetup{
  linkcolor=blue,
  urlcolor=blue,
  citecolor=blue,
  filecolor=blue,
  colorlinks=true
}

\allowdisplaybreaks

\newcommand{\beq}{\begin{equation}}
\newcommand{\eeq}{\end{equation}}

\journal{Physics Letters B}

\biboptions{sort&compress}

\begin{document}

\begin{frontmatter}

\title{Exploring the Truth and Beauty of Theory Landscapes with Machine Learning}

\author[UF]{Konstantin T.~Matchev\fnref{contribution}} 
\author[UF]{Katia Matcheva\fnref{contribution}}
\author[UF]{Pierre Ramond\fnref{contribution}}
\author[UF]{Sarunas~Verner\fnref{contribution}}

\fntext[contribution]{All authors share equal contributions to this paper.}
\affiliation[UF]{organization={Institute for Fundamental Theory, Physics Department, University of Florida},
            city={Gainesville},
            state={FL},
            postcode={32611}, 
            country={USA}}

\begin{abstract}
Theoretical physicists describe nature by i) building a theory model and ii) determining the model parameters. The latter step involves the dual aspect of both fitting to the existing experimental data and satisfying abstract criteria like beauty, naturalness, etc. We use the Yukawa quark sector as a toy example to demonstrate how both of those tasks can be accomplished with machine learning techniques. We propose loss functions whose minimization results in true models that are also beautiful as measured by three different criteria --- uniformity, sparsity, or symmetry.
\end{abstract}

\end{frontmatter}
    
\section{Introduction}
The task of a theoretical physicist is to develop a theory model describing a set of natural physics phenomena. The first step, often requiring great insight and inspiration, is to choose a theoretical framework, which inevitably depends on a certain set of input parameters $\{P_i\}$, $i=1,2,\ldots, N_P$. The second step is more straightforward, and involves determining the numerical values of these parameters which ``were chosen by nature". There are two aspects of this parameter-fitting process:
\begin{itemize}
\item {\em Truth.} In order to be viable, a model has to be above all truthful. We take this to mean that the model correctly accounts for the existing measurements of all independent experimental observables $\{O_\alpha\}$, $\alpha=1,2,\ldots, N_O$, sensitive to the parameters $\{P_i\}$. This is accomplished by tuning the model parameters $\{P_i\}$ until the model predictions fit the data. This adjustment can typically be accomplished successfully, since in most cases the number of tunable model parameters exceeds the number of available independent measurements, i.e., $N_P> N_O$. 
\item {\em Beauty.} After this fitting procedure, we are typically still left with $N_P - N_O$ undetermined model parameters.\footnote{Some of the remaining parameters could perhaps be probed in future experiments by measuring additional observables. The most common example of this sort is given by the masses and couplings of particles yet to be discovered.} Those remaining degrees of freedom can then be adjusted in order to make the model more ``beautiful". However, beauty is in the eye of the beholder, and there is no single universally accepted notion of ``beauty" in theoretical physics. To this day, such concepts as fine-tuning, naturalness, simplicity, etc., continue to be hotly debated in the theoretical community. 
\end{itemize}

However, from a machine learning standpoint there is nothing mysterious about ``beauty", as long as it can be quantified, i.e., one adopts an agreed-upon, community-wide quantitative measure indicating the ``beauty" of a model. In the past, a number of such measures have been introduced to quantify the fine-tuning in new physics models like low-energy supersymmetry \citep{Barbieri:1987fn,Anderson:1994dz,Anderson:1994tr,Feng:1999mn,Feng:1999zg}. Once a quantitative measure of the model's beauty is adopted, the fitting of the model parameters becomes a simple optimization problem amenable to machine learning approaches.

\begin{figure}
  \centering
    \includegraphics[height=0.5\textwidth]{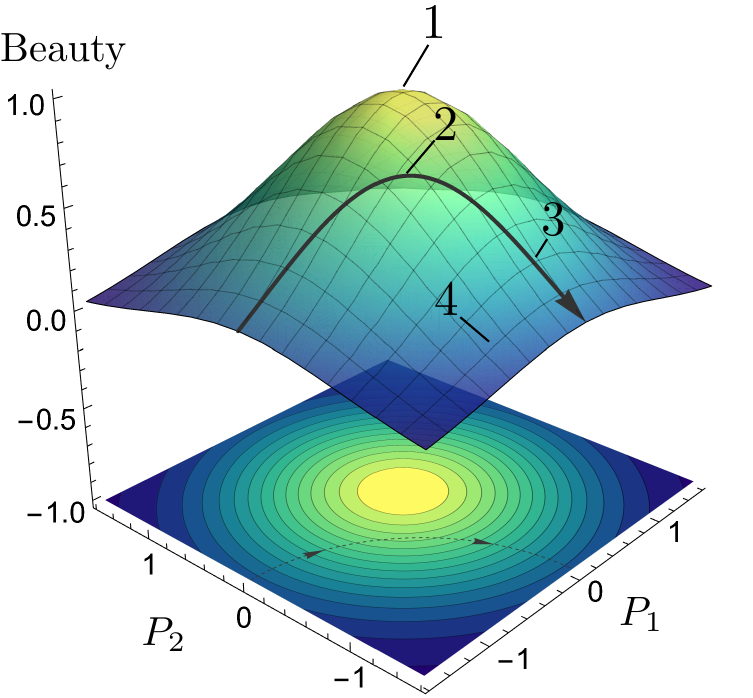}
  \caption{A schematic depiction of the quantitative measure of ``beauty" of a theory model as a function of the model parameters $P_1$ and $P_2$. In the absence of any experimental data, model ``1'' is the most beautiful of all. An experimental measurement of a single observable $O$ enforces a constraint (shown with the dashed line) among the model parameters. Model ``2" is the most beautiful model which is consistent with experiment. Model ``3" is an example of a true and not-as-beautiful model, while model ``4" is an example of a model which is neither true nor beautiful. }
  \label{fig:landscape}
\end{figure}

The procedure described above is schematically illustrated in Fig.~\ref{fig:landscape}, which shows a plot of the quantitative measure of ``beauty" (vertical axis) versus the model parameters (for ease of illustration, we choose $N_P=2$). In the absence of any experimental data, the theorists would agree that model ``1'' is ``the fairest of them all". However, experimental data typically ends up modifying this naive conclusion. For example, consider the measurement of a single observable (i.e., the case of $N_O=1$), which enforces a constraint among the model parameters. This constraint is schematically depicted with the dashed line in the horizontal plane, which in turn defines a path (marked with the solid line) along the ``beauty surface". Models along this path are ``true", and model ``2" is the most beautiful one among them. Model ``3" is an example of a true and not-as-beautiful model, while model ``4" is an example of a model which is neither true nor beautiful.

Operationally, the optimization needed to arrive at model ``2" can be performed in three different ways.
\begin{itemize}
\item {\em Constrained optimization.} The experimental constraints are incorporated into the beauty loss function (e.g., with Lagrange multipliers) and the optimization is done over the full domain of the parameters $P_i$.
\item {\em Unconstrained optimization.} Alternatively, one may try to solve all the constraints explicitly. In the language of Fig.~\ref{fig:landscape}, this means deriving an analytical parametrization of the dashed line in the horizontal plane, then optimizing the beauty loss function along this line.
\item {\em Hybrid optimization.} If not all constraints can be easily solved, one can use a hybrid approach, where some constraints are solved explicitly while others are added to the loss. This will be our approach here.
\end{itemize}

In this paper, we focus on the flavor sector, which is arguably the ``ugliest" part of the Standard Model. We consider several possible choices for quantitative measures of the beauty of the model. In each case, we define corresponding loss functions whose minimization by construction yields ``the most truthful and beautiful" model. Preliminary results from this work were reported at the 37th Conference on Neural Information Processing Systems (NeurIPS) \cite{Matchev:2023mii}; here we have added the additional constraint of CP-violation and a third illustrative example (described in Section~\ref{subsec:symmetric}).

Our approach should be viewed as part of a much broader program of trying to learn the laws of nature with a machine, eliminating any human intervention whatsoever \cite{Langley1977, Langley1987,Kokar1986,Langley1989,Zembowicz1992,Todorovski1997,Bongard2007,Schmidt2009,Battaglia2016,Chang2016,Guimera2020}. For example, it has been demonstrated that the machine can re-derive the known classical physics laws from data \cite{Udrescu:2019mnk,Cranmer:2020wew,liu2022ai,https://doi.org/10.48550/arxiv.2206.10540}. Symbolic learning was recently successfully applied to problems in a wide range of physics areas, e.g., 
in astrophysics \cite{Cranmer2019,Cranmer:2020wew,Delgado:2021cuw}, 
in astronomy for the study of orbital dynamics \cite{Iten2020,Lemos:2022cdj}
and exoplanet transmission spectroscopy  \cite{Matchev2022ApJ},
in collider physics \cite{Choi:2010wa,Butter:2021rvz,Dersy:2022bym,Alnuqaydan:2022ncd,Dong:2022trn}, 
in materials science \cite{wang_wagner_rondinelli_2019},
and in behavioral science \cite{Arechiga2021}.
Our approach is slightly less ambitious than those studies, since we already adopt the mathematical framework for the description of the phenomena, and instead focus only on the determination of the ``best" model parameters which, within that mathematical framework, might have been chosen by nature. Along similar lines, to find the allowed U(1) charge assignments in models with a U(1) flavor symmetry, Ref.~\cite{Nishimura:2020nre} used reinforcement learning, as opposed to an exhaustive computerized search as in \cite{Lee:2007fw,Lee:2007qx}.

The paper is organized as follows. In Section~\ref{sec:SMparams}, we introduce our notation and the experimental inputs to our subsequent numerical analysis. Then in Section~\ref{sec:quarksector}, we consider three examples of ``beautiful" quark textures. First, in Section~\ref{subsec:democratic}, we consider beauty to mean uniformity, i.e., the elements in the Yukawa matrices have the same magnitude. Then in Section~\ref{subsec:heterogeneous}, we take beauty to mean sparsity, i.e., the Yukawa matrices have a large number of vanishing elements. Finally, in Section~\ref{subsec:symmetric}, we identify beauty with the symmetry (in the sense of linear algebra) of the Yukawa matrices. Section~\ref{sec:conclusions} contains our summary and conclusions.

\section{Standard Model Parameters}
\label{sec:SMparams}
In our analysis, we primarily use the notation and conventions from the standard textbook by M. Schwartz~\cite{Schwartz:2014sze}. The Lagrangian governing the quark mass sector can be expressed as
\begin{equation}
    \mathcal{L}_{\rm quarks} \; = \; -Y_{ij}^d \bar{Q}^i H d_{R}^j-Y_{ij}^u \bar{Q}^i \widetilde{H} u_R^j+ \rm{h.c.} \, ,
    \label{eq:quarksec}
\end{equation}
where $Q^i$, with $i=1,2,3$, represents the quark doublets of the $SU(2)_L$ group, incorporating three generations
\begin{equation}
Q^i \; = \; \left(\begin{array}{c}
u_{L}^i \\
d_{L}^i
\end{array}\right) \, ,
\end{equation}
$H$ denotes the Higgs field, and $\widetilde{H}$ is its conjugate field, defined as
\begin{equation}
\widetilde{H} \; \equiv \; i \sigma_2 H^\ast \, .
\end{equation} 
Here $\sigma_2$ is the second Pauli matrix and the asterisk indicates complex conjugation. Thus, we have,
\begin{equation}
H \; = \; \left(\begin{array}{c}
H^{+} \\
H^0
\end{array}\right)\, , 
\quad 
\widetilde{H} \; = \; 
\left(\begin{array}{c}
H^{0 *} \\
-H^{-}
\end{array}\right) \, .
\end{equation}
The fields $u_R^i$ and $d_R^i$ are the right-handed $SU(2)_L$ quark singlets, given by
\begin{equation}
u_R^i \; = \; \{u_R, c_R, t_R \}, \qquad d_R^i \; = \; \{d_R, s_R, b_R \} \, .
\end{equation}
This formulation incorporates all three quark generations. The Yukawa matrices $Y_{ij}^u$ and $Y_{ij}^d$ are arbitrary complex matrices and are not constrained by hermiticity or similar mathematical properties. As complex $3 \times 3$ matrices, they have a total of $18$ degrees of freedom. After spontaneous symmetry breaking, the neutral Higgs field component, $H^0$, acquires a vacuum expectation value of $v/\sqrt{2}$, modifying the quark mass Lagrangian (\ref{eq:quarksec}) to
\begin{equation}
    \mathcal{L}_{\rm quarks} \; = \; - \frac{v}{\sqrt{2}} \left[\bar{d}_L Y^d d_R + \bar{u}_L Y^u u_R \right] + \rm{h.c.}
    \label{eq:quarksec2}
\end{equation}
In the interaction eigenstate basis, the quark mass matrices are not diagonal, but the interactions between the up- and down-type quarks and $W^\pm$-bosons are. In this basis, the quark mass matrices are given by 
\begin{equation}
(M_u)_{ij} \; \equiv \; \frac{v}{\sqrt{2}} Y^u_{ij} \,, \qquad
(M_d)_{ij} \; \equiv \; \frac{v}{\sqrt{2}} Y^d_{ij} \, .
\label{eq:mass_matrices}
\end{equation}
Diagonalization of these matrices is achieved by transitioning to a new mass-eigenstate basis
\begin{subequations}
\begin{eqnarray}
u'_L &= U_u u_L \,, \qquad d'_L &= U_d d_L \, ,\\
u'_R &= K_u u_R \,, \qquad d'_R &= K_d d_R \, ,
\end{eqnarray}
\end{subequations}
where the primed quark fields are mass eigenstates, and $U_u$, $U_d$, $K_u$, and $K_d$ are unitary rotation matrices. The resulting diagonal mass matrices in this new basis are
\begin{equation}
M'_u \; = \; U_u\, M_u\, K_u^\dagger \, , \qquad
M'_d \; = \; U_d\, M_d\, K_d^\dagger \, .
\label{eq:mass_matrices_diagonal}
\end{equation}
The unitary matrices $U_u$ and $K_u$ are derived by diagonalizing the Hermitian matrices $M_uM_u^\dagger$ and $M_u^\dagger M_u$, respectively. Given the Hermitian nature of these matrices, they each possess three eigenvalues associated with the squared masses of the up-type quarks, namely the up quark, charm quark, and top quark, with respective masses $m^u_i = \{m_u, m_c, m_t\}$. In parallel, the matrices $U_d$ and $K_d$ are determined through the diagonalization of $M_dM_d^\dagger$ and $M_d^\dagger M_d$. This process yields the masses for the down-type quarks, specifically the down quark, strange quark, and bottom quark, represented as $m^d_i = \{m_d, m_s, m_b\}$.

In the mass eigenstate basis, the quark mass Lagrangian assumes a diagonal form and can be expressed as:
\begin{equation}
    \mathcal{L}_{\rm quarks} \; = \; 
    - m_i^u \delta_{ij} \bar{u}'{}_L^i u'{}_R^j 
    - m_i^d \delta_{ij} {\bar{d}}'{}_L^i d'{}_R^j 
    + \rm{h.c.} 
    \label{eq:quarksec4}
\end{equation}
However, quark mixing phenomena become evident in the interactions that involve charged $W$ bosons, typically parameterized using the Cabibbo-Kobayashi-Maskawa (CKM) matrix:
\begin{equation}
V_{\rm CKM} \; \equiv \; U_u^{\dagger} U_d=\left(\begin{array}{lll}
V_{11} & V_{12} & V_{13} \\
V_{21} & V_{22} & V_{23} \\
V_{31} & V_{32} & V_{33}
\end{array}\right) \; = \; \left(\begin{array}{ccc}
V_{u d} & V_{u s} & V_{u b} \\
V_{c d} & V_{c s} & V_{c b} \\
V_{t d} & V_{t s} & V_{t b}
\end{array}\right) \, .
\end{equation}
The CKM matrix is a $3 \times 3$ complex unitary matrix with nine degrees of freedom. Utilizing the $U(1)$ symmetry of the $6$ quark mass terms in the Lagrangian~(\ref{eq:quarksec4}), $5$ degrees of freedom can be removed, leaving $4$ independent parameters.\footnote{It is not possible to remove all $6$ phase degrees of freedom through $U(1)$ rotations because these rotations require distinct phases to alter the CKM matrix. If the phases in these rotations were identical, the CKM matrix would remain unaffected.} These are commonly parametrized as 
\small 
\begin{align}
V_{\rm CKM}= & \left(\begin{array}{ccc}
1 & 0 & 0 \\
0 & \cos \theta_{23} & \sin \theta_{23} \\
0 & -\sin \theta_{23} & \cos \theta_{23}
\end{array}\right) \left(\begin{array}{ccc}
\cos \theta_{13} & 0 & \sin \theta_{13} e^{i \delta} \\
0 & 1 & 0 \\
-\sin \theta_{13} e^{i \delta} & 0 & \cos \theta_{13}
\end{array}\right)\left(\begin{array}{ccc}
\cos \theta_{12} & \sin \theta_{12} & 0 \\
-\sin \theta_{12} & \cos \theta_{12} & 0 \\
0 & 0 & 1
\end{array}\right) \nonumber \\
= & \left(\begin{array}{ccc}
c_{12} c_{13} & s_{12} c_{13} & s_{13} e^{-i \delta} \\
-s_{12} c_{23}-c_{12} s_{23} s_{13} e^{i \delta} & c_{12} c_{23}-s_{12} s_{23} s_{13} e^{i \delta} & s_{23} c_{13} \\
s_{12} s_{23}-c_{12} c_{23} s_{13} e^{i \delta} & -c_{12} s_{23}-s_{12} c_{23} s_{13} e^{i \delta} & c_{23} c_{13}
\end{array}\right) \, ,
\end{align}
\normalsize
where $s_{ij} \equiv \sin{\theta_{ij}}$ and $c_{ij} \equiv \cos{\theta_{ij}}$, and $\delta$ is the CP-violating phase. Assuming unitarity of the CKM matrix, the matrix parameters are given by~\cite{ParticleDataGroup:2022pth}
\begin{align}
\sin \theta_{12} & =0.22500 \pm 0.00067 \,, & \sin \theta_{13} & =0.00369 \pm 0.00011 \, , \nonumber \\
\sin \theta_{23} & =0.04182_{-0.00074}^{+0.00085} \, , & \delta & =1.144 \pm 0.027 \, . \nonumber
\end{align} 
The fit results for the magnitudes of all nine CKM elements are
\begin{equation}
\begin{aligned}
\left|V_{\rm {CKM, exp} }\right| & =\left(\begin{array}{lll}
0.97435 \pm 0.00016 & 0.22500 \pm 0.00067 & 0.00369 \pm 0.00011 \\
0.22486 \pm 0.00067 & 0.97349 \pm 0.00016 & 0.04182_{-0.000074}^{+0.00085} \\
0.00857_{-0.00018}^{+0.00020} & 0.0410_{-0.00072}^{+0.0003} & 0.999118_{-0.000036}^{+0.0000031}
\end{array}\right) \, .
\end{aligned}
\label{eq:CKMexp}
\end{equation}
The measurement of CP violation, independent of phase convention, is given by the Jarlskog invariant, defined as
\begin{equation}
    J \; = \;  \Im \left[V_{i j} V_{k l} V_{i l}^* V_{k j}^*\right]=J \sum_{m, n} \varepsilon_{i k m} \varepsilon_{j l n} \, ,
\end{equation}
and its experimental value is given by~\cite{ParticleDataGroup:2022pth}
\begin{equation}   
    \label{eq:jarlskogexp}
    J \; = \; \left(3.08^{+0.15}_{-0.13} \right) \times 10^{-5} \, .
\end{equation}

In addition to Eq. (\ref{eq:CKMexp}), our analysis also incorporates the running quark masses at a chosen reference energy scale, with the top quark mass scale being the preferred choice~\cite{Babu:2009fd} (although other scales like the $Z$ mass scale are also viable~\cite{Giraldo:2018mqi}). These values, along with their experimental uncertainties, are summarized in Table~\ref{table:masses}.

\begin{table}
  \caption{Quark masses (with uncertainties) evaluated at the top quark mass scale.}
  \label{table:masses}
  \centering
  \begin{tabular}{cccccc}
    \toprule
    $m_u~(\rm{MeV})$  & $m_d~(\rm{MeV})$ & $m_c~(\rm{GeV})$ & $m_s~(\rm{MeV})$ & $m_b~(\rm{GeV})$ & $m_t~(\rm{GeV})$  \\
    \toprule
    $1.22_{-0.15}^{+0.28} $ & $2.76_{-0.10}^{+0.28} $ & $ 0.59_{-0.01}^{+0.01}$ & $52_{-1.89}^{+4.79}$ & $2.75_{-0.01}^{+0.02}$& $162.9_{-0.28}^{+0.28}$\\  
    \bottomrule
  \end{tabular}
\end{table}

\section{Quark Sector Textures}
\label{sec:quarksector}
In this section, we construct our loss functions and demonstrate their utility through three examples. Following \cite{Matchev:2023mii}, the initial inputs to the Lagrangian (\ref{eq:quarksec}) are the two Yukawa matrices $Y^u$ and $Y^d$, or equivalently, the mass matrices $M_u$ and $M_d$ given in eq.~(\ref{eq:mass_matrices}). These matrices encompass a total of $36$ degrees of freedom. Out of these, $15$ ($9$ from the CKM matrix and $6$ from the quark masses) are determined by the experimental inputs in Eq.~(\ref{eq:CKMexp}) and Table~\ref{table:masses}. Thus, we could theoretically set this as an optimization problem in a $36$-dimensional space, constrained by $15$ parameters. However, to expedite the optimization process, we instead consider a parametrization of $M_u$ and $M_d$ that inherently satisfies the quark mass constraints. This is achieved through the inverse relations to equations~(\ref{eq:mass_matrices_diagonal}):
\begin{equation}
M_u = U_u^\dagger\, M'_u\, K_u, \qquad
M_d = U_d^\dagger\, M'_d\, K_d.
\label{eq:mass_matrices_diagonal_inverse}
\end{equation}
By setting the diagonal elements in the matrices $M'_u$ and $M'_d$ to magnitudes equal to the respective quark masses, we automatically fulfill the quark mass constraints. This approach reduces the degrees of freedom in each matrix $M_u$ and $M_d$ to 15 (that matches exactly the degrees of freedom between the left- and the right-hand sides of the equations~(\ref{eq:mass_matrices_diagonal_inverse}).\footnote{We note that in our approach, we do not manually fix any degrees of freedom; instead, we allow the minimization of the loss function to determine all the degrees of freedom.} Consequently, we have effectively transformed the problem into an optimization task in a $30$-dimensional space, subject to only the 9 constraints detailed in (\ref{eq:CKMexp}). To ensure these constraints are met, we define the following loss function:
\begin{equation}
L_{\rm CKM} \; = \; \sum_{ij} \left(|V_{\rm CKM}|_{ij} - |V_{\rm{CKM, exp}}|_{ij}\right)^2 \, .
\label{eq:loss_CKM}
\end{equation}

To account for the CP-violating phase, we introduce a loss function that incorporates the Jarlskog invariant, which is a phase-convention-independent measure of CP violation:
\begin{equation}   
    L_{\rm Jarlskog} \; = \; \left( \left| \Im \left[V_{us} V_{cb} V_{ub}^* V_{cs}^* \right] \right| - J_{\rm exp} \right)^2 \, .
    \label{eq:loss_Jarlskog}
\end{equation}

For each of the three examples discussed below, we will employ a series of pseudo-experiments for illustration. In every pseudo-experiment, our starting point are the experimental values 
specified in Eqs.~(\ref{eq:CKMexp}) and (\ref{eq:jarlskogexp}) and in Table~\ref{table:masses}. 
However, in order to account for the experimental uncertainties in the measurements, in our numerical analysis we sample the experimental inputs from split normal distributions whose left (right) standard deviations match the lower (upper) experimental uncertainties of the respective experimental values. Moving forward, our results will be expressed in terms of the mass matrices $M_u$ and $M_d$, as opposed to using the dimensionless Yukawa matrices $Y^u$ and $Y^d$.

\subsection{Uniform Texture}
\label{subsec:democratic}
The fundamental nature and origin of the Yukawa matrices $Y^u$ and $Y^d$ remains one of the key unsolved mysteries within the Standard Model (SM), marking the ``flavor problem" as a vibrant focus of theoretical research for over half a century. Numerous theories, often featuring novel symmetries, particles, and interactions, have been proposed to solve these ``Yukawa textures." Experimental outcomes will ultimately determine the validity of these models by either confirming or disproving these additional structures and components. Our approach here adopts a bottom-up methodology within the SM, treated as an effective theory, where the only available experimental data are those in Eqs.~(\ref{eq:CKMexp}) and (\ref{eq:jarlskogexp}), and Table~\ref{table:masses}. Thus, the primary criterion for selecting one model over another hinges on whether the resulting Yukawa sector embodies a certain ``beauty" or not.

For an initial exploration, let's define a ``beautiful" flavor model as one that promotes uniformity, meaning that each element within a Yukawa matrix has approximately the same magnitude. We express this as
\begin{equation}
   | Y^u_{i j} | \; \simeq \; | Y^u_{k l} |\, , \qquad \forall i,j,k,l \, .
\end{equation}
We can impose this uniformity criterion by introducing the following loss function
\begin{equation}
    \label{eq:lossuniform_up}
    L_{\rm const, up} \; = \; \sum_{i,j,k,l}  \left(|Y^u_{ij}| - |Y^u_{kl}| \right)^2 \, .
\end{equation}
Likewise, applying uniformity to the down-type Yukawa matrix leads to
\begin{equation}
   | Y^d_{i j} | \; \simeq \; | Y^d_{k l} | \, , \qquad \forall i,j,k,l \, ,
\end{equation}
with the corresponding loss function
\begin{equation}
    \label{eq:lossuniform_down}
    L_{\rm const, down} \; = \;  \sum_{i,j,k,l} \left(|Y^d_{ij}| - |Y^d_{kl}| \right)^2 \, .
\end{equation}

\begin{figure}
  \centering
    \includegraphics[height=0.5\textwidth]{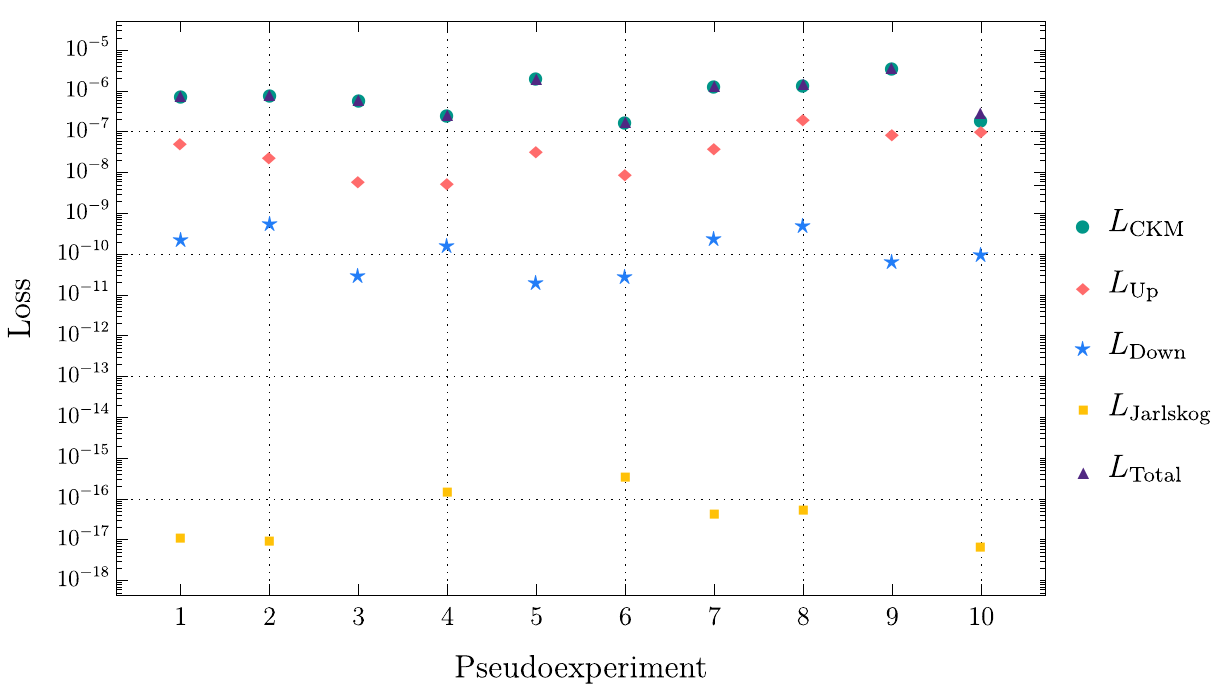}
  \caption{The trained total loss values (see Eq. \ref{eq:lossuniform}) along with the detailed breakdown into the four individual components (Eqs. \ref{eq:loss_CKM}, \ref{eq:loss_Jarlskog}, \ref{eq:lossuniform_up}, and \ref{eq:lossuniform_down}) are presented for ten representative pseudo-experiments. These experiments are part of the uniform texture analysis discussed in Section \ref{subsec:democratic}.}
  \label{fig:lossuniform}
\end{figure}

Thus, the complete loss function for uniform Yukawa textures can be expressed as
\begin{equation}
    \label{eq:lossuniform}
    L_{\rm uniform} \; = \; L_{\rm CKM} +  \frac{1}{m_t}L_{\rm const, up} +  \frac{1}{m_b} L_{\rm const, down} + \frac{1}{J_{\rm exp}^2} L_{\rm Jarlskog} \, .
\end{equation}

In this equation, the loss functions $L_{\rm const, up}$ and $L_{\rm const, down}$ are normalized with respect to the masses of the heaviest quarks, ensuring that all three contributions to the loss function are equally weighted. Similarly, $L_{\rm Jalrskog}$, is normalized with respect to $J_{\rm exp}^2$.

We conduct $10$ pseudo-experiments, focusing on minimizing the full loss function~(\ref{eq:lossuniform}). The results are illustrated in Fig.~\ref{fig:lossuniform}. Here, we present not only the final values for the overall loss achieved through training but also enumerate the individual contributions from Eqs.~(\ref{eq:loss_CKM}), (\ref{eq:loss_Jarlskog}), (\ref{eq:lossuniform_up}), and (\ref{eq:lossuniform_down}). Across each pseudo-experiment, we consistently observed minimal loss values, signifying the potential viability of the Yukawa textures under consideration.

To provide a concrete example, we quote a specific outcome from one of the $10$ pseudo-experiments (noting that the results from the others were similar). For the up-type quark mass matrix, we obtain
\begin{equation}
 M_u \; = \;
\begin{pmatrix}
52.1055 - 14.8303 i & -51.2066 + 17.6863 i & 50.9441 + 18.4286 i \\
41.4531 + 34.8792 i & -42.9766 - 32.9838 i & 13.8632 + 52.3711 i \\
-50.9028 + 18.5424 i & 49.9240 - 21.0360 i & -52.2362 - 14.3631 i
\end{pmatrix} \, ,
\label{Mu_uniform}
\end{equation}
and
\begin{equation}
 |M_u| \; = \;
\begin{pmatrix}
54.1749 &  54.1749 & 54.1749  \\
54.1749 & 54.1749 & 54.1749 \\
54.1749& 54.1749 & 54.1749
\end{pmatrix} \, .
\end{equation}
For the down-type mass matrix $M_d$, we find
\begin{equation}
 M_d \; = \;
\begin{pmatrix}
-0.3248 + 0.8549i & -0.6992 + 0.5895i & 0.9090 - 0.1008i \\
-0.1510 - 0.9020i & 0.3066 - 0.8616i & -0.6748 + 0.6173i \\
-0.2776 + 0.8714i & -0.6623 + 0.6307i & 0.8913 - 0.2048i\\
\end{pmatrix} \, ,
\label{Md_uniform}
\end{equation}
and
\begin{equation}
 |M_d| \; = \;
\begin{pmatrix}
0.9145 & 0.9145 & 0.9146 \\
0.9146 & 0.9145 & 0.9146 \\
0.9145 & 0.9146 & 0.9145
\end{pmatrix} \, .
\end{equation}

\begin{figure}[t]
  \centering
    \includegraphics[height=0.4\textwidth]{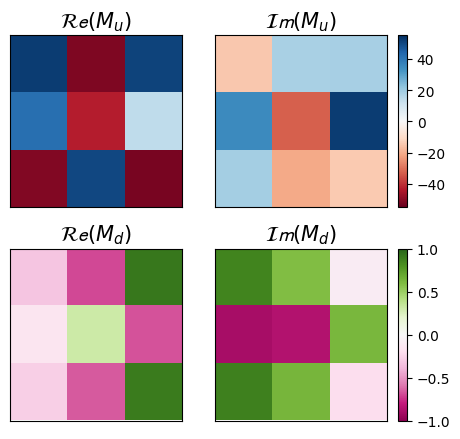}  ~~~
    \includegraphics[height=0.4\textwidth]{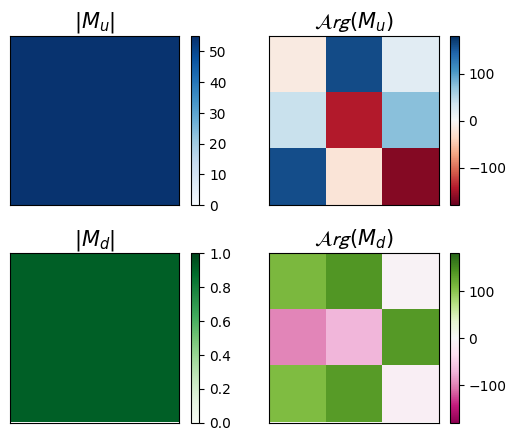}      
  \caption{The learned mass matrices $M_u$ (upper panels) and $M_d$ (lower panels), corresponding to the uniform Yukawa texture case discussed in Sec.~\ref{subsec:democratic}. Each panel illustrates a specific learned matrix, with the color bar denoting the values of the individual elements of the matrix.}
  \label{fig:uniformplots}
\end{figure}

The results of this pseudo-experiment are illustrated in Fig.~\ref{fig:uniformplots}. The top panels of the figure relate to Eq.~(\ref{Mu_uniform}) and the bottom panels to Eq.~(\ref{Md_uniform}). Each row is divided into four panels: the first two on the left display the real and imaginary parts of the matrix elements, while the last two panels exhibit their magnitude and phase. Notably, in each mass matrix, the magnitudes of the various elements are approximately equal, aligning well with our predefined criterion of ``beauty" for this example.

\subsection{Zero Textures}
\label{subsec:heterogeneous}

\begin{table}[b]
  \caption{A number $N$ of vanishing elements in the mass matrix (top row), 
  the total number of patterns (middle row), and the number of potentially acceptable patterns (bottom row).}
  \label{table:patterns}
  \centering
  \begin{tabular}{ccccccccc}
    \toprule
    N        & 1 & 2  & 3 & 4 & 5 & 6 & 7 & 8 \\
    \toprule
    All patterns & 9 & 36 & 84 & 126 & 126 & 84 & 36 & 9 \\    \midrule
    Acceptable  & 9 & 36 & 78 &  81 &  36 &  6 &  0 & 0 \\ 
    \bottomrule
  \end{tabular}
\end{table}

In our second example, we explore the concept of zero textures, as discussed in \cite{Ramond:1993kv}, where the ``beauty" of a model is measured by its sparsity. This concept aims to maximize the number of vanishing elements in the Yukawa matrices. In our analysis, we treat the number of such vanishing elements, $N$, as a variable hyperparameter, as detailed in Table~\ref{table:patterns}. For each chosen value of $N$, we are free to set to zero $N$ elements in the matrix. The total number of possible patterns is indicated in the second row of Table~\ref{table:patterns}. However, some patterns can be immediately ruled out as they lead to at least one zero mass eigenvalue. The count of the remaining, potentially feasible, patterns is presented in the bottom row of the table.

\begin{figure}[t]
  \centering
    \includegraphics[height=0.4\textwidth]{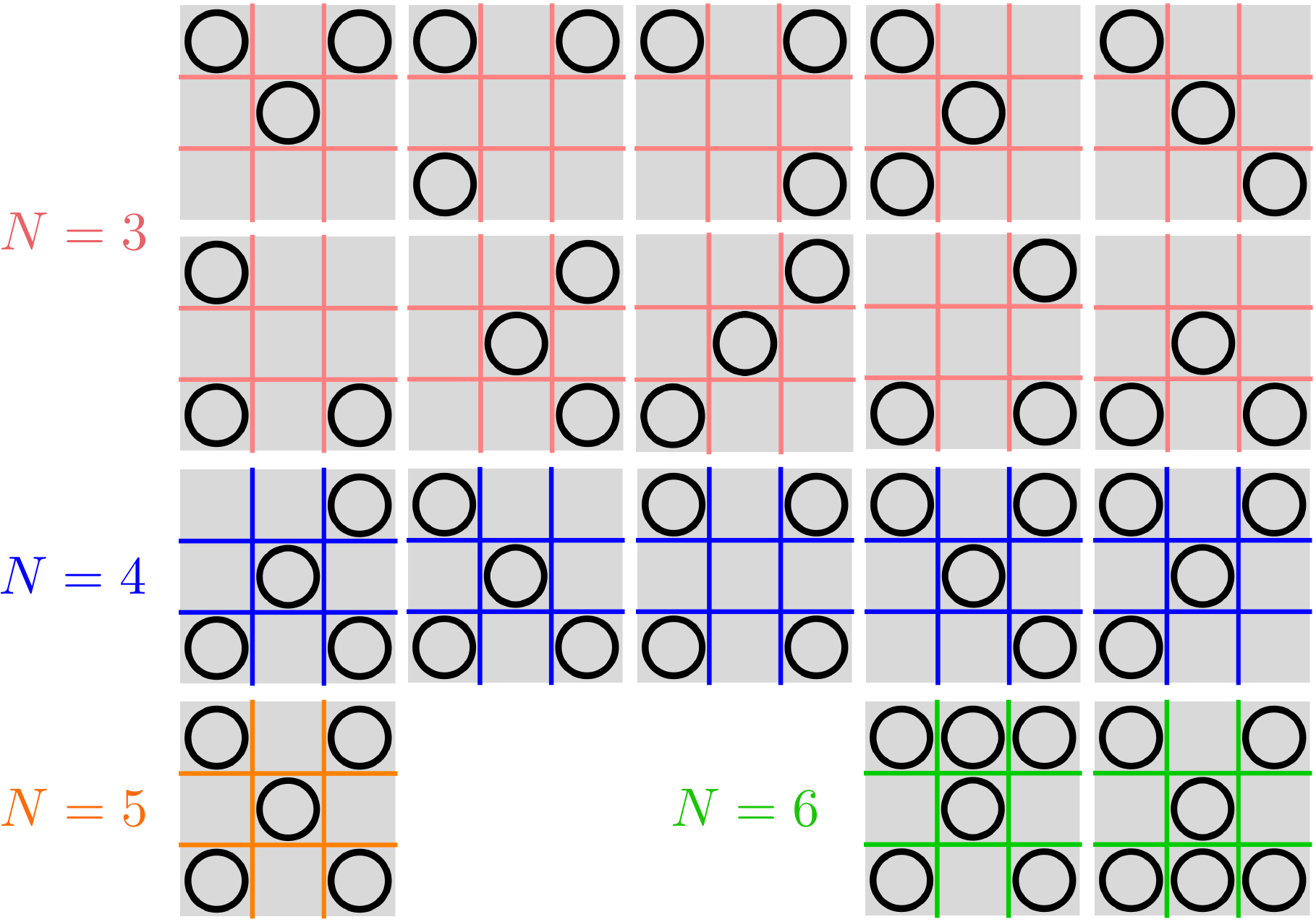}
  \caption{The zero texture patterns examined in the example of Sec.~\ref{subsec:heterogeneous}, with circles marking the matrix elements set to zero.}
  \label{fig:patterns}
\end{figure}

For this study, we focus on several representative patterns shown in Fig.~\ref{fig:patterns}, where circles indicate the positions of the matrix elements set to zero. Let ${\cal S}$ denote the set of zero locations in a specific pattern, for example, ${\cal S}={11,22,13}$ for the first $N=3$ pattern in Fig.~\ref{fig:patterns}. The corresponding loss functions for these patterns are defined as follows
\begin{equation}
    L_{\rm zeros, up} \; = \; 
    \sum_{{ij}\in {\cal S}} |Y^u_{ij}|^2 \, ,
\end{equation}
and 
\begin{equation}
    L_{\rm zeros, down} \; = \; 
    \sum_{{ij}\in {\cal S}} |Y^d_{ij}|^2 \, .
\end{equation}
We then minimize the loss function that includes the following components:
\begin{equation}
    L \; = \; L_{\rm CKM} + \frac{1}{m_u} L_{\rm zeros, up} +  \frac{1}{m_d} L_{\rm zeros, down} + \frac{1}{J_{\rm exp}^2} L_{\rm Jarlskog} \, .
\end{equation}
We note that for increased precision in the minimization process, we normalize the loss functions $L_{\rm zeros, up}$ and $L_{\rm zeros, down}$
with respect to the masses of the lightest quarks. 

\begin{figure}[t]
  \centering
    \includegraphics[height=0.5\textwidth]{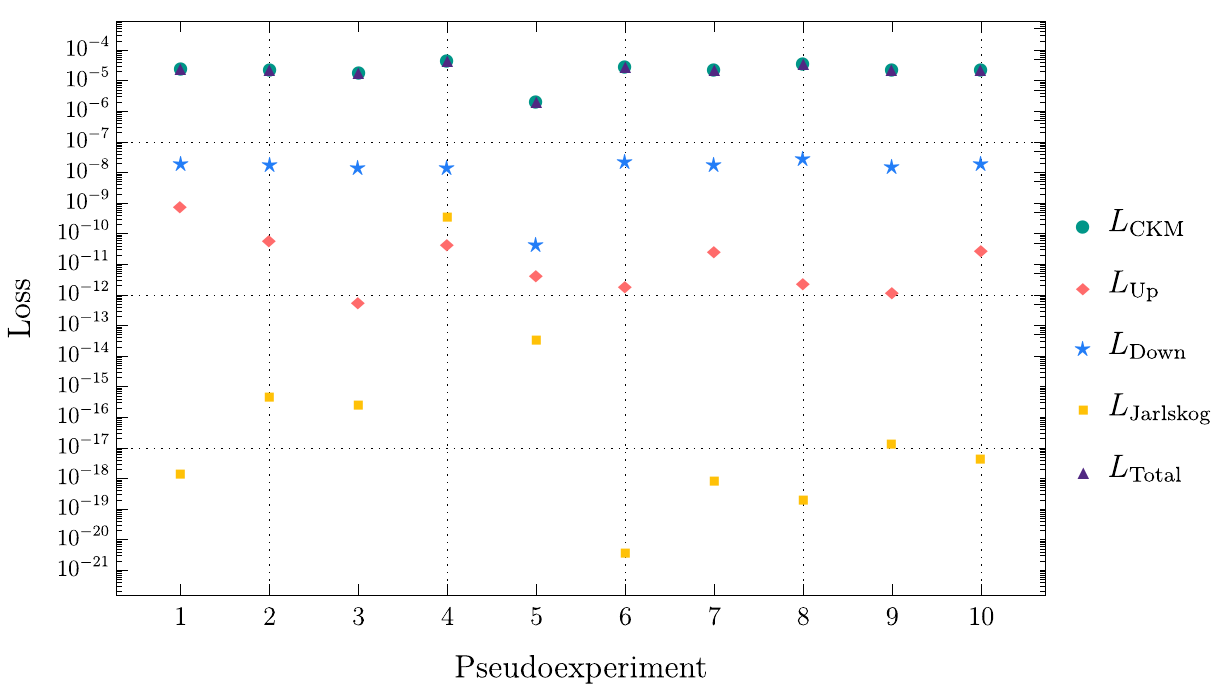}
  \caption{As in Fig.~\ref{fig:lossuniform}, it illustrates the trained loss values and their components for the ten $N=3$ zero texture patterns from Fig.~\ref{fig:patterns}, averaged across 10 different pseudo-experiments.}
  \label{fig:loss_zeros}
\end{figure}

\begin{figure}[t]
  \centering
    \includegraphics[height=0.40\textwidth]{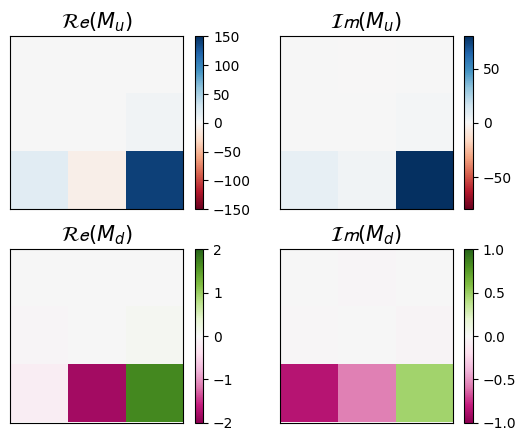}  ~~~
    \includegraphics[height=0.40\textwidth]{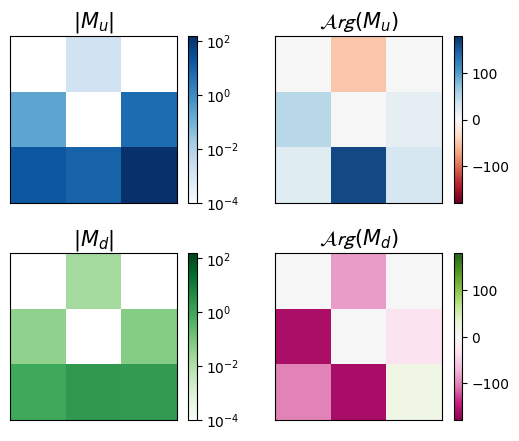}      
  \caption{ As in Fig.~\ref{fig:uniformplots}, but for the three zero texture result from Eqs.~(\ref{Mu_zero}) and (\ref{Md_zero}).}
  \label{fig:M_3zeros}
\end{figure}

Our findings indicate that feasible patterns lead to low loss values. In the same manner as Fig.~\ref{fig:lossuniform}, Fig.~\ref{fig:loss_zeros} presents the trained loss values and their components for the ten $N=3$ zero texture patterns, averaged over 10 pseudo-experiments. As expected, all $N=3$ zero texture patterns are plausible. The result of one such pseudo-experiment for the first pattern in Fig.~\ref{fig:patterns} is given by
\begin{equation}
\label{Mu_zero}
   M_u \; = \; \left(
\begin{array}{ccc}
0.0000 - 0.0000i & 0.0010 - 0.0012i & 0.0000 + 0.0000i \\
0.1446 + 0.1709i & 0.0000 + 0.0000i & 5.4710 + 1.7690i \\
17.0333 + 6.8434i& -8.9194 + 2.9354i & 140.3701 + 79.7760i \\
\end{array}
\right) \, ,
\end{equation}
\begin{equation}
\label{Mu_zero2}
   |M_u| \; = \; \left(
\begin{array}{ccc}
0.0000 & 0.0016 & 0.0000 \\
0.2239 & 0.0000 & 5.7499 \\
18.3566& 9.3900 & 161.4558 \\
\end{array}
\right) \, ,
\end{equation}
and 
\begin{equation}
\label{Md_zero}
   M_d \; = \; \left(
\begin{array}{ccc}
0.0000 + 0.0000 i & 0.0009 - 0.0177i & 0.0000 + 0.0000i \\
-0.0385 - 0.0137i & 0.0000- 0.0000 i & 0.0555 - 0.0311i \\
-0.1596 - 0.8464i & -1.8296 - 0.5630i & 1.6871 + 0.4743i \\
\end{array}
\right) \, ,
\end{equation}
\begin{equation}
\label{Md_zero2}
   |M_d| \; = \; \left(
\begin{array}{ccc}
0.0000 & 0.0177 &  0.0000 \\
0.0409 & 0.0000 & 0.0636\\
0.8613 & 1.9143 & 1.7525 \\
\end{array}
\right) \, .
\end{equation}
This result is shown in Fig.~\ref{fig:M_3zeros}, confirming that the matrix entries at positions 11, 22, and 13 are notably small.

\begin{figure}[t]
  \centering
    \includegraphics[height=0.35\textwidth]{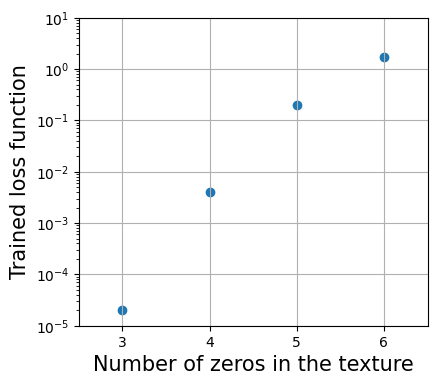}
  \caption{
  Averaged values of the trained loss as a function of the number of texture zeros.  }
  \label{fig:Nzeros}
\end{figure}

The outcomes for larger values of $N$ are summarized in Fig.~\ref{fig:Nzeros}. Here, we present the computed loss values, averaged across both the number of pseudo-experiments ($10$ in this case) and the different patterns from Fig.~\ref{fig:patterns} for each respective $N$ value. 

\subsection{Symmetric Textures}
\label{subsec:symmetric}
In our final example, we consider symmetric Yukawa textures where the off-diagonal elements are ``mirrored" (in magnitude) across the main diagonal:
\begin{equation}
    |Y_{ij}^u| \; = \; |Y_{ji}^u| \, , \qquad \forall i,j\, .
\end{equation}
This condition can be imposed by introducing the following loss function
\begin{equation}
    L_{\rm sym, up} \; = \; \sum_{i,j \in \mathcal{S}} \left(|Y_{ij}^u| - |Y_{ji}^u| \right)^2 \, .
\end{equation}
Similarly, for the down-type Yukawa matrix, the analogous condition is
\begin{equation}
    |Y_{ij}^d| \; = \; |Y_{ji}^d| \, , \qquad \forall i,j \, .
\end{equation}
and the corresponding loss function is given by
\begin{equation}
    L_{\rm sym, down} \; = \; \sum_{i,j \in \mathcal{S}} \left(|Y_{ij}^d| - |Y_{ji}^d| \right)^2 \, .
\end{equation}
The complete loss function for symmetric Yukawa textures is given by
\begin{equation}
    L \; = \; L_{\rm CKM} + \frac{1}{m_t}L_{\rm sym, up} + \frac{1}{m_b}L_{\rm sym, down} + \frac{1}{J_{\rm exp}^2} L_{\rm Jarlskog} \, .
\end{equation}

\begin{figure}[ht!]
  \centering
    \includegraphics[height=0.5\textwidth]{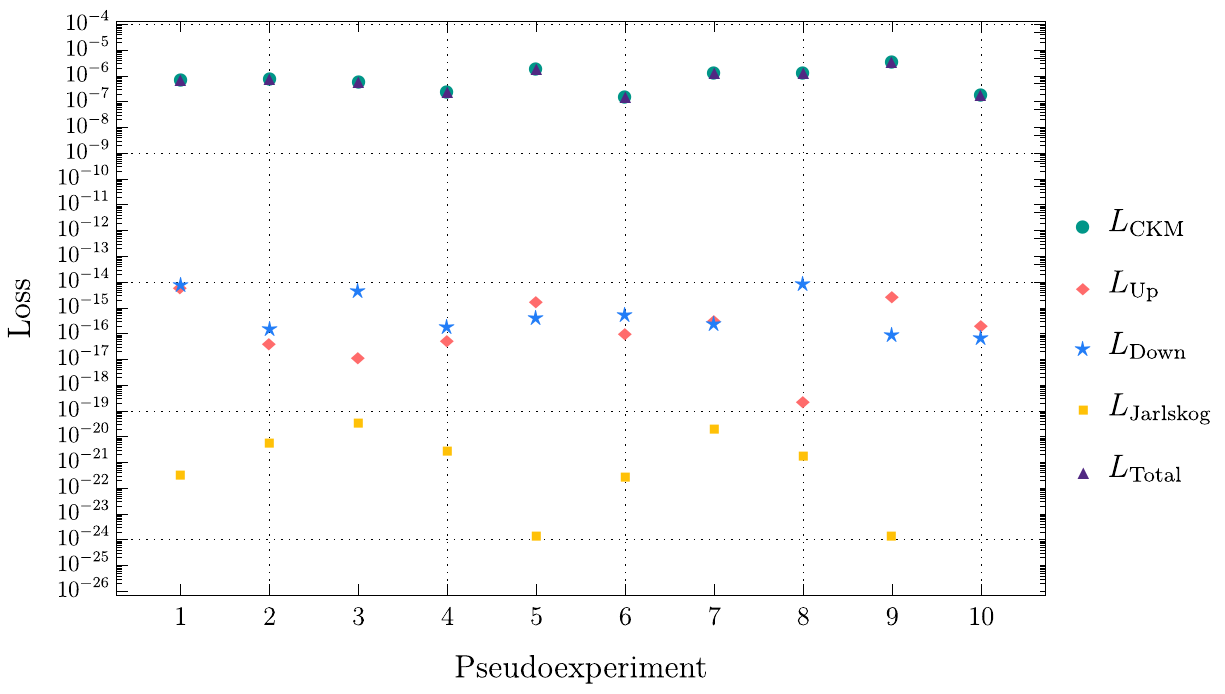}
  \caption{As in Fig.~\ref{fig:lossuniform}, it illustrates the trained loss values and their components for the 10 symmetric texture patterns, averaged across 10 different pseudo-experiments.}
  \label{fig:losssym}
\end{figure}

\begin{figure}[ht!]
  \centering
    \includegraphics[height=0.40\textwidth]{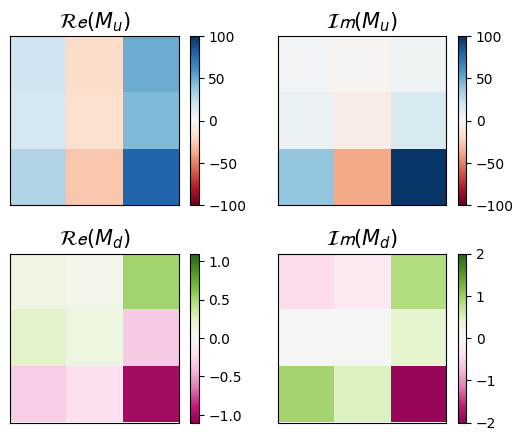}  ~~~
    \includegraphics[height=0.40\textwidth]{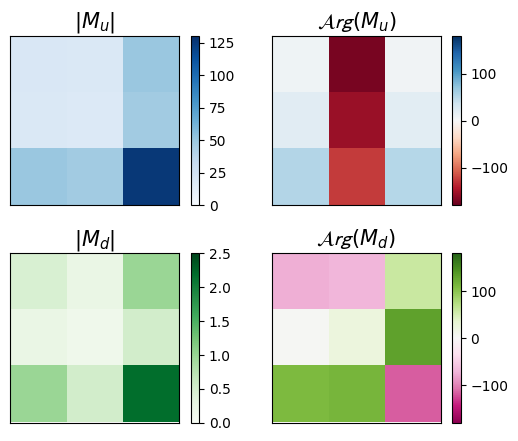}      
  \caption{ As in Fig.~\ref{fig:uniformplots}, but with the symmetric texture result from Eqs.~(\ref{Mu_symmetric}) and (\ref{Md_symmetric}).}
  \label{fig:M_symmetric}
\end{figure}

Similarly to Figs.~\ref{fig:lossuniform} and \ref{fig:loss_zeros}, Fig.~\ref{fig:losssym} shows (the individual contributions to) the trained loss values in 10 representative pseudo-experiments. For one randomly chosen pseudo-experiment, we find the up-type quark mass matrix
\begin{equation}
 M_u \; = \;
\begin{pmatrix}
19.5575 + 2.5557i & -18.2841 - 2.8153i & 49.5441 + 5.0476i \\
17.3188 + 6.5033i & -16.1554 - 7.2142i & 44.1967 + 15.1457i \\
30.2621 + 39.5513i & -27.2614 - 37.9415i &79.2378 + 97.6495i\\
\end{pmatrix} \, ,
\label{Mu_symmetric}
\end{equation}
with
\begin{equation}
 |M_u| \; = \;
\begin{pmatrix}
19.7238 & 18.4996 & 49.8006\\
18.4996 & 17.6930 & 46.7198 \\
49.8006 & 46.7198 & 125.7540\\
\end{pmatrix} \, ,
\label{eq:absMusym}
\end{equation}
and the down-type mass matrix
\begin{equation}
 M_d \; = \;
\begin{pmatrix}
0.1123 - 0.4304i & 0.0777 - 0.2201i & 0.5203 + 0.8316i\\
0.2331 + 0.0127i & 0.1303 + 0.0548i & -0.3273 + 0.3973i \\
-0.3106 + 0.9305i &-0.2084 + 0.4707i & -1.0108 - 1.9185i\\
\end{pmatrix} \, ,
\label{Md_symmetric}
\end{equation}
with
\begin{equation}
 |M_d| \; = \;
\begin{pmatrix}
0.4448 & 0.2334 & 0.9810 \\
0.2334 & 0.1414 & 0.5148 \\
0.9810 & 0.5148 & 2.1685 \\
\end{pmatrix} \, .
\label{eq:absMdsym}
\end{equation}
As seen in Eqs.~(\ref{eq:absMusym}) and (\ref{eq:absMdsym}), the obtained matrices are perfectly symmetric in the sense defined above. Fig.~\ref{fig:M_symmetric} is a pictorial illustration of the result displayed in eqs.~(\ref{Mu_symmetric}) and (\ref{Md_symmetric}).

\section{Summary and Conclusions}
\label{sec:conclusions}

In theoretical particle physics, the process of developing new models entails meeting the objective constraints of the existing experimental data, as well as subjective criteria like beauty and naturalness set forth by the theoretical physics community. To achieve both of these goals, we employ machine learning techniques with suitably designed loss functions addressing the perceived deficiencies in the Yukawa sector of the Standard Model. With the three toy examples from Section~\ref{sec:quarksector}, we showed that this approach yields models that are not only consistent with the experimental data, but also possess the desired aesthetic elegance as defined by a quantitative benchmark. In future work, we plan to extend this analysis to the lepton sector of the Standard Model as well.

{\bf Acknowledgements.}
We thank K.~Babu, S.~Gleyzer, D.~Gon\c{c}alves, K.~Kong and G.~Shiu for useful discussions. This work is supported in part by the U.S.~Department of Energy award number DE-SC0022148.

\bibliography{references}

\begin{thebibliography}{10}
\expandafter\ifx\csname url\endcsname\relax
  \def\url#1{\texttt{#1}}\fi
\expandafter\ifx\csname urlprefix\endcsname\relax\def\urlprefix{URL }\fi
\expandafter\ifx\csname href\endcsname\relax
  \def\href#1#2{#2} \def\path#1{#1}\fi

\bibitem{Barbieri:1987fn}
R.~Barbieri, G.~F. Giudice, {Upper Bounds on Supersymmetric Particle Masses}, Nucl. Phys. B 306 (1988) 63--76.
\newblock \href {https://doi.org/10.1016/0550-3213(88)90171-X} {\path{doi:10.1016/0550-3213(88)90171-X}}.

\bibitem{Anderson:1994dz}
G.~W. Anderson, D.~J. Castano, {Measures of fine tuning}, Phys. Lett. B 347 (1995) 300--308.
\newblock \href {http://arxiv.org/abs/hep-ph/9409419} {\path{arXiv:hep-ph/9409419}}, \href {https://doi.org/10.1016/0370-2693(95)00051-L} {\path{doi:10.1016/0370-2693(95)00051-L}}.

\bibitem{Anderson:1994tr}
G.~W. Anderson, D.~J. Castano, {Naturalness and superpartner masses or when to give up on weak scale supersymmetry}, Phys. Rev. D 52 (1995) 1693--1700.
\newblock \href {http://arxiv.org/abs/hep-ph/9412322} {\path{arXiv:hep-ph/9412322}}, \href {https://doi.org/10.1103/PhysRevD.52.1693} {\path{doi:10.1103/PhysRevD.52.1693}}.

\bibitem{Feng:1999mn}
J.~L. Feng, K.~T. Matchev, T.~Moroi, {Multi - TeV scalars are natural in minimal supergravity}, Phys. Rev. Lett. 84 (2000) 2322--2325.
\newblock \href {http://arxiv.org/abs/hep-ph/9908309} {\path{arXiv:hep-ph/9908309}}, \href {https://doi.org/10.1103/PhysRevLett.84.2322} {\path{doi:10.1103/PhysRevLett.84.2322}}.

\bibitem{Feng:1999zg}
J.~L. Feng, K.~T. Matchev, T.~Moroi, {Focus points and naturalness in supersymmetry}, Phys. Rev. D 61 (2000) 075005.
\newblock \href {http://arxiv.org/abs/hep-ph/9909334} {\path{arXiv:hep-ph/9909334}}, \href {https://doi.org/10.1103/PhysRevD.61.075005} {\path{doi:10.1103/PhysRevD.61.075005}}.

\bibitem{Matchev:2023mii}
K.~T. Matchev, K.~Matcheva, P.~Ramond, S.~Verner, {Seeking Truth and Beauty in Flavor Physics with Machine Learning}, in: {37th Conference on Neural Information Processing Systems}, 2023.
\newblock \href {http://arxiv.org/abs/2311.00087} {\path{arXiv:2311.00087}}.

\bibitem{Langley1977}
P.~Langley, Bacon: A production system that discovers empirical laws, in: IJCAI, 1977.

\bibitem{Langley1987}
P.~Langley, H.~A. Simon, G.~L. Bradshaw, \href{https://doi.org/10.1007/978-3-642-82742-6_2}{Heuristics for Empirical Discovery}, Springer Berlin Heidelberg, Berlin, Heidelberg, 1987, pp. 21--54.
\newblock \href {https://doi.org/10.1007/978-3-642-82742-6_2} {\path{doi:10.1007/978-3-642-82742-6_2}}.
\newline\urlprefix\url{https://doi.org/10.1007/978-3-642-82742-6_2}

\bibitem{Kokar1986}
M.~Kokar, Determining arguments of invariant functional descriptions., Machine Learning 1 (1986) 403--422.
\newblock \href {https://doi.org/10.1023/A:1022818816206} {\path{doi:10.1023/A:1022818816206}}.

\bibitem{Langley1989}
P.~Langley, J.~M. Zytkow, \href{https://www.sciencedirect.com/science/article/pii/0004370289900519}{Data-driven approaches to empirical discovery}, Artificial Intelligence 40~(1) (1989) 283--312.
\newblock \href {https://doi.org/https://doi.org/10.1016/0004-3702(89)90051-9} {\path{doi:https://doi.org/10.1016/0004-3702(89)90051-9}}.
\newline\urlprefix\url{https://www.sciencedirect.com/science/article/pii/0004370289900519}

\bibitem{Zembowicz1992}
R.~Zembowicz, J.~M. \.{Z}ytkow, Discovery of equations: Experimental evaluation of convergence, in: Proceedings of the Tenth National Conference on Artificial Intelligence, AAAI'92, AAAI Press, 1992, p. 70–75.

\bibitem{Todorovski1997}
L.~Todorovski, S.~Dzeroski, Declarative bias in equation discovery, in: Proceedings of the Fourteenth International Conference on Machine Learning, Morgan Kaufmann, 1997, pp. 376--384.

\bibitem{Bongard2007}
J.~{Bongard}, H.~{Lipson}, {From the Cover: Automated reverse engineering of nonlinear dynamical systems}, Proceedings of the National Academy of Science 104~(24) (2007) 9943--9948.
\newblock \href {https://doi.org/10.1073/pnas.0609476104} {\path{doi:10.1073/pnas.0609476104}}.

\bibitem{Schmidt2009}
M.~{Schmidt}, H.~{Lipson}, {Distilling Free-Form Natural Laws from Experimental Data}, Science 324~(5923) (2009) 81.
\newblock \href {https://doi.org/10.1126/science.1165893} {\path{doi:10.1126/science.1165893}}.

\bibitem{Battaglia2016}
P.~W. {Battaglia}, R.~{Pascanu}, M.~{Lai}, D.~{Rezende}, K.~{Kavukcuoglu}, {Interaction Networks for Learning about Objects, Relations and Physics}, arXiv e-prints (2016) arXiv:1612.00222\href {http://arxiv.org/abs/1612.00222} {\path{arXiv:1612.00222}}.

\bibitem{Chang2016}
M.~B. {Chang}, T.~{Ullman}, A.~{Torralba}, J.~B. {Tenenbaum}, {A Compositional Object-Based Approach to Learning Physical Dynamics}, arXiv e-prints (2016) arXiv:1612.00341\href {http://arxiv.org/abs/1612.00341} {\path{arXiv:1612.00341}}.

\bibitem{Guimera2020}
R.~Guimerà, I.~Reichardt, A.~Aguilar-Mogas, F.~A. Massucci, M.~Miranda, J.~Pallarès, M.~Sales-Pardo, \href{https://www.science.org/doi/abs/10.1126/sciadv.aav6971}{A bayesian machine scientist to aid in the solution of challenging scientific problems}, Science Advances 6~(5) (2020) eaav6971.
\newblock \href {http://arxiv.org/abs/https://www.science.org/doi/pdf/10.1126/sciadv.aav6971} {\path{arXiv:https://www.science.org/doi/pdf/10.1126/sciadv.aav6971}}, \href {https://doi.org/10.1126/sciadv.aav6971} {\path{doi:10.1126/sciadv.aav6971}}.
\newline\urlprefix\url{https://www.science.org/doi/abs/10.1126/sciadv.aav6971}

\bibitem{Udrescu:2019mnk}
S.-M. Udrescu, M.~Tegmark, {AI Feynman: a Physics-Inspired Method for Symbolic Regression}, Sci. Adv. 6~(16) (2020) eaay2631.
\newblock \href {http://arxiv.org/abs/1905.11481} {\path{arXiv:1905.11481}}, \href {https://doi.org/10.1126/sciadv.aay2631} {\path{doi:10.1126/sciadv.aay2631}}.

\bibitem{Cranmer:2020wew}
M.~Cranmer, A.~Sanchez-Gonzalez, P.~Battaglia, R.~Xu, K.~Cranmer, D.~Spergel, S.~Ho, {Discovering Symbolic Models from Deep Learning with Inductive Biases} (6 2020).
\newblock \href {http://arxiv.org/abs/2006.11287} {\path{arXiv:2006.11287}}.

\bibitem{liu2022ai}
Z.~Liu, V.~Madhavan, M.~Tegmark, \href{https://link.aps.org/doi/10.1103/PhysRevE.106.045307}{Machine learning conservation laws from differential equations}, Phys. Rev. E 106 (2022) 045307.
\newblock \href {https://doi.org/10.1103/PhysRevE.106.045307} {\path{doi:10.1103/PhysRevE.106.045307}}.
\newline\urlprefix\url{https://link.aps.org/doi/10.1103/PhysRevE.106.045307}

\bibitem{https://doi.org/10.48550/arxiv.2206.10540}
Y.~Matsubara, N.~Chiba, R.~Igarashi, T.~Taniai, Y.~Ushiku, \href{https://arxiv.org/abs/2206.10540}{Rethinking symbolic regression datasets and benchmarks for scientific discovery} (2022).
\newblock \href {http://arxiv.org/abs/2206.10540} {\path{arXiv:2206.10540}}, \href {https://doi.org/10.48550/ARXIV.2206.10540} {\path{doi:10.48550/ARXIV.2206.10540}}.
\newline\urlprefix\url{https://arxiv.org/abs/2206.10540}

\bibitem{Cranmer2019}
M.~D. Cranmer, R.~Xu, P.~Battaglia, S.~Ho, \href{https://arxiv.org/abs/1909.05862}{Learning symbolic physics with graph networks} (2019).
\newblock \href {http://arxiv.org/abs/1909.05862} {\path{arXiv:1909.05862}}, \href {https://doi.org/10.48550/ARXIV.1909.05862} {\path{doi:10.48550/ARXIV.1909.05862}}.
\newline\urlprefix\url{https://arxiv.org/abs/1909.05862}

\bibitem{Delgado:2021cuw}
A.~M. Delgado, D.~Wadekar, B.~Hadzhiyska, S.~Bose, L.~Hernquist, S.~Ho, {Modelling the galaxy\textendash{}halo connection with machine learning}, Mon. Not. Roy. Astron. Soc. 515~(2) (2022) 2733--2746.
\newblock \href {http://arxiv.org/abs/2111.02422} {\path{arXiv:2111.02422}}, \href {https://doi.org/10.1093/mnras/stac1951} {\path{doi:10.1093/mnras/stac1951}}.

\bibitem{Iten2020}
R.~{Iten}, T.~{Metger}, H.~{Wilming}, L.~{del Rio}, R.~{Renner}, {Discovering Physical Concepts with Neural Networks}, Physical Review Letters 124~(1) (2020) 010508.
\newblock \href {http://arxiv.org/abs/1807.10300} {\path{arXiv:1807.10300}}, \href {https://doi.org/10.1103/PhysRevLett.124.010508} {\path{doi:10.1103/PhysRevLett.124.010508}}.

\bibitem{Lemos:2022cdj}
P.~Lemos, N.~Jeffrey, M.~Cranmer, S.~Ho, P.~Battaglia, {Rediscovering orbital mechanics with machine learning}, Mach. Learn. Sci. Tech. 4~(4) (2023) 045002.
\newblock \href {http://arxiv.org/abs/2202.02306} {\path{arXiv:2202.02306}}, \href {https://doi.org/10.1088/2632-2153/acfa63} {\path{doi:10.1088/2632-2153/acfa63}}.

\bibitem{Matchev2022ApJ}
K.~T. {Matchev}, K.~{Matcheva}, A.~{Roman}, {Analytical Modeling of Exoplanet Transit Spectroscopy with Dimensional Analysis and Symbolic Regression}, The Astrophysical Journal 930~(1) (2022) 33.
\newblock \href {http://arxiv.org/abs/2112.11600} {\path{arXiv:2112.11600}}, \href {https://doi.org/10.3847/1538-4357/ac610c} {\path{doi:10.3847/1538-4357/ac610c}}.

\bibitem{Choi:2010wa}
S.~Choi, {Construction of a Kinematic Variable Sensitive to the Mass of the Standard Model Higgs Boson in $H \to WW^* \to l^+ \nu l^- \bar{\nu}$ using Symbolic Regression}, JHEP 08 (2011) 110.
\newblock \href {http://arxiv.org/abs/1006.4998} {\path{arXiv:1006.4998}}, \href {https://doi.org/10.1007/JHEP08(2011)110} {\path{doi:10.1007/JHEP08(2011)110}}.

\bibitem{Butter:2021rvz}
A.~Butter, T.~Plehn, N.~Soybelman, J.~Brehmer, {Back to the Formula -- LHC Edition} (9 2021).
\newblock \href {http://arxiv.org/abs/2109.10414} {\path{arXiv:2109.10414}}.

\bibitem{Dersy:2022bym}
A.~Dersy, M.~D. Schwartz, X.~Zhang, {Simplifying Polylogarithms with Machine Learning} (6 2022).
\newblock \href {http://arxiv.org/abs/2206.04115} {\path{arXiv:2206.04115}}.

\bibitem{Alnuqaydan:2022ncd}
A.~Alnuqaydan, S.~Gleyzer, H.~Prosper, {SYMBA: symbolic computation of squared amplitudes in high energy physics with machine learning}, Mach. Learn. Sci. Tech. 4~(1) (2023) 015007.
\newblock \href {http://arxiv.org/abs/2206.08901} {\path{arXiv:2206.08901}}, \href {https://doi.org/10.1088/2632-2153/acb2b2} {\path{doi:10.1088/2632-2153/acb2b2}}.

\bibitem{Dong:2022trn}
Z.~Dong, K.~Kong, K.~T. Matchev, K.~Matcheva, {Is the machine smarter than the theorist: Deriving formulas for particle kinematics with symbolic regression}, Phys. Rev. D 107~(5) (2023) 055018.
\newblock \href {http://arxiv.org/abs/2211.08420} {\path{arXiv:2211.08420}}, \href {https://doi.org/10.1103/PhysRevD.107.055018} {\path{doi:10.1103/PhysRevD.107.055018}}.

\bibitem{wang_wagner_rondinelli_2019}
Y.~Wang, N.~Wagner, J.~M. Rondinelli, Symbolic regression in materials science, MRS Communications 9~(3) (2019) 793–805.
\newblock \href {https://doi.org/10.1557/mrc.2019.85} {\path{doi:10.1557/mrc.2019.85}}.

\bibitem{Arechiga2021}
N.~Arechiga, F.~Chen, Y.-Y. Chen, Y.~Zhang, R.~Iliev, H.~Toyoda, K.~Lyons, Accelerating understanding of scientific experiments with end to end symbolic regression (2021).
\newblock \href {http://arxiv.org/abs/2112.04023} {\path{arXiv:2112.04023}}.

\bibitem{Nishimura:2020nre}
S.~Nishimura, C.~Miyao, H.~Otsuka, {Exploring the flavor structure of quarks and leptons with reinforcement learning}, JHEP 23 (2020) 021.
\newblock \href {http://arxiv.org/abs/2304.14176} {\path{arXiv:2304.14176}}, \href {https://doi.org/10.1007/JHEP12(2023)021} {\path{doi:10.1007/JHEP12(2023)021}}.

\bibitem{Lee:2007fw}
H.-S. Lee, K.~T. Matchev, T.~T. Wang, {A U(1) -prime solution to the $\mu^-$ problem and the proton decay problem in supersymmetry without R-parity}, Phys. Rev. D 77 (2008) 015016.
\newblock \href {http://arxiv.org/abs/0709.0763} {\path{arXiv:0709.0763}}, \href {https://doi.org/10.1103/PhysRevD.77.015016} {\path{doi:10.1103/PhysRevD.77.015016}}.

\bibitem{Lee:2007qx}
H.-S. Lee, C.~Luhn, K.~T. Matchev, {Discrete gauge symmetries and proton stability in the U(1)-prime - extended MSSM}, JHEP 07 (2008) 065.
\newblock \href {http://arxiv.org/abs/0712.3505} {\path{arXiv:0712.3505}}, \href {https://doi.org/10.1088/1126-6708/2008/07/065} {\path{doi:10.1088/1126-6708/2008/07/065}}.

\bibitem{Schwartz:2014sze}
M.~D. Schwartz, {Quantum Field Theory and the Standard Model}, Cambridge University Press, 2014.

\bibitem{ParticleDataGroup:2022pth}
R.~L. Workman, et~al., {Review of Particle Physics}, PTEP 2022 (2022) 083C01.
\newblock \href {https://doi.org/10.1093/ptep/ptac097} {\path{doi:10.1093/ptep/ptac097}}.

\bibitem{Babu:2009fd}
K.~S. Babu, {TASI Lectures on Flavor Physics}, in: {Theoretical Advanced Study Institute in Elementary Particle Physics}: {The Dawn of the LHC Era}, 2010, pp. 49--123.
\newblock \href {http://arxiv.org/abs/0910.2948} {\path{arXiv:0910.2948}}, \href {https://doi.org/10.1142/9789812838360_0002} {\path{doi:10.1142/9789812838360_0002}}.

\bibitem{Giraldo:2018mqi}
Y.~Giraldo, E.~Rojas, {Five Non-Fritzsch Texture Zeros for Quarks Mass Matrices in the Standard Model}, in: {38th International Symposium on Physics in Collision}, 2018.
\newblock \href {http://arxiv.org/abs/1811.05068} {\path{arXiv:1811.05068}}.

\bibitem{Ramond:1993kv}
P.~Ramond, R.~G. Roberts, G.~G. Ross, {Stitching the Yukawa quilt}, Nucl. Phys. B 406 (1993) 19--42.
\newblock \href {http://arxiv.org/abs/hep-ph/9303320} {\path{arXiv:hep-ph/9303320}}, \href {https://doi.org/10.1016/0550-3213(93)90159-M} {\path{doi:10.1016/0550-3213(93)90159-M}}.

\end{thebibliography}

\bibliographystyle{elsarticle-num}

\end{document}